\documentstyle[12pt]{article}

\newcommand{\newsection}{    
\setcounter{equation}{0}
\section}
\newcommand{\tr}[1]{\,{\rm tr}\,#1\,}
\def\e{{\,\rm e}\,}

\def\eop{\vspace*{\fill}\pagebreak}
\def\be{\begin{equation}}
\def\ee{\end{equation}}
\def\bea{\begin{eqnarray}}
\def\eea{\end{eqnarray}}

\newcommand{\rf}[1]{(\ref{#1})}
\newcommand{\eq}[1]{Eq.~(\ref{#1})}

\def\bb{\bar{\beta}}

\def\l{\lambda}
\def\h{\eta}

\newcommand{\ie}{{\it i.e.}\ }

\newcommand{\ra}{\rightarrow}
\newcommand{\fr}[2]{{\textstyle {#1 \over #2}}}

\begin{document}

\begin{flushright}
ITEP-YM-5-92 \\ July, 1992
\end{flushright}

\begin{center}
{\LARGE The Problem of Large-N Phase Transition \\ \vspace{0.6cm} in
Kazakov--Migdal Model of Induced QCD} \end{center} \vspace{.5cm}
\begin{center}
{\large S.\ Khokhlachev} \\
\mbox{} \\ {\it Cybernetics Council, Academy of Science} \\
{\it Vavilov st. 40,  117333 Moscow, Russia} \\
\vspace{0.5cm} \mbox{} \\ {\large and} \\
\vspace{0.5cm} \mbox{} \\
{\large Yu.\ Makeenko}\footnote{E--mail: \ makeenko@itep.msk.su \ \ / \ \
makeenko@desyvax.bitnet \ \ / \ \ makeenko@nbivax.nbi.dk } \\ \mbox{} \\
{\it Institute of Theoretical and Experimental Physics} \\
{\it B.Cheremuskinskaya 25, 117259 Moscow, Russia}
\end{center}

\vspace{1cm}

\begin{abstract}
We study  the lattice gauge model proposed recently by Kazakov and
Migdal for inducing QCD. We discuss an extra local $Z_N$
which is a symmetry of the model and propose of how to construct
observables. We discuss the role of the large-$N$ phase transition
which should occur before the one associated with the continuum limit
in order that the model describes continuum QCD.
We formulate the mean field approach to study the large-$N$ phase transition
for an arbitrary potential and show that no first order phase transition
occurs for the quadratic potential.
\end{abstract}


\eop

\newsection {Introduction}

Recently Kazakov and Migdal \cite{KM92} have proposed a very interesting
lattice gauge model for inducing QCD. The model is defined by the partition
function
\be
Z=\int \prod_{x,\mu} dU_{\mu}(x) \prod_x d\Phi(x)
\e^{\sum_x N \tr{\left(-V[\Phi(x)]+
\sum_{\mu}\Phi(x)U_\mu(x)\Phi(x+\mu)U_\mu^\dagger(x)\right)}}
\label{partition}
\ee
where the field $\Phi(x)$ takes values in the adjoint representation of the
gauge group $SU(N)$ and the link variable $U_\mu(x)$ belongs to the group.
The potential $V[\Phi]$ can be expanded at small lattice spacing $a$ in the
power series
\be
V[\Phi]= m_0^2 \Phi^2+\lambda \Phi^4 + \ldots
\label{potential}
\ee
with $m_0$ being the bare mass of the field $\Phi$.
A strong coupling solution to this model has been constructed by
Migdal~\cite{Mig92a} at $N=\infty$ and then extended to next orders of the
$1/N$-expansion \cite{Mig92b}.

To obtain the solution two assumptions have been made. The first one is that
the critical point $m_0^2=D$ ($D$ is the dimensionality of euclidean
space-time), which separates the strong coupling region, is associated
with continuum QCD. The second one is that the path integral over $\Phi(x)$
is saturated by a single $x$-independent configuration --- the master field.

On our mind along solving the Kazakov--Migdal model, one should better
understand qualitative properties of the model like how to construct
observables or how to recover the area law in the continuum. As usual
the question of which phase transitions occur in the lattice system is
of special importance. The knowledge in advance of the phase structure
might help to understand what kind of the distribution of eigenvalues
of the matrices $\Phi_{ij}(x)$ would correspond to a physical solution.
In this paper we apply a standard lattice gauge theory technique to study
phase transitions in the Kazakov--Migdal model.

A subtle point with the model \rf{partition} is that it possesses an extra
local $Z_N$ symmetry
\be
U_\mu(x) \ra Z_\mu(x) U_\mu(x) .
\label{symmetry}
\ee
The important role of this symmetry in the Kazakov--Migdal model has been
pointed out recently by Kogan, Semenoff and Weiss~\cite{KSW92}.
These authors argued that due to the $Z_N$-symmetry the model undergoes a phase
transition which was assumed to occur simultaneously with the transition from
strong to weak coupling solution. This phase transition separates two phases
which
differ by the nature of confinement --- the so-called local confinement
versus usual area law.

The main subject of the present paper is a study of properties of the
Kazakov--Migdal model which come from the presence of the $Z_N$-symmetry.
To calculate observables, in particularly the string tension,
we propose a general procedure based on the Wilson loops in the adjoint
representation $SU(N)$ which are invariant under the $Z_N$-symmetry and
can be expressed entirely in terms of averages over $\Phi(x)$.
We show that, being proper normalized
to be of order 1 in the naive continuum limit, these quantities are
$\sim 1/N^2$ in the strong coupling expansion.
For the quadratic potential $V[\Phi]$, we look for the
corresponding critical point at $N=\infty$ using the mean field method.
We show  that no phase transition occurs in $D=4$ up to the point
$m_0^2=D$ where the model becomes unstable.

We discuss that, if the large-$N$ phase transition occurs before the one
associated with the continuum limit, a consistent picture of inducing
continuum QCD by the Kazakov--Migdal model emerges.
In this case one gets below the large-$N$ phase transition a theory
with usual confinement (area law) rather than local confinement as
in each order of the strong coupling expansion.
We formulate the mean field approach to study the large-$N$ phase transition
for the case of an arbitrary $V[\Phi]$.

We speculate that the phase transition associated with the continuum limit
should be identified with the one separating confinement and Higgs phases
and discuss the general scenario of inducing QCD by the Kazakov--Migdal model.
We point out that the mean field analyses of the Higgs phase transition is
reduced to a one-matrix problem which is solvable for cubic and logarithmic
potentials.

\newsection{$Z_N$ and observables}

The local $Z_N$ symmetry%
\footnote{Since the transformation can be done independently on each link of
the lattice, such a symmetry was called in Ref.\cite{KhM81} that of the 3-rd
kind.}
has far-reaching consequences for the model \rf{partition}. First of all,
the average of any non-invariant quantity like the Wilson loop in the
fundamental representation vanishes except for a loop passing the same contour
back and forth (\ie with vanishing minimal area). Moreover, this property
holds independently of how many phase transitions the system undergoes under
the way to the continuum since the local symmetry can not be broken
spontaneously. 
Therefore, the average of the {\it fundamental} Wilson loop always vanishes.

To obtain the continuum Wilson loops from the Kazakov--Migdal model, we
propose to consider averages of the {\it adjoint} Wilson loops
\be
W_A(C)=\left\langle\frac{1}{N^2}
\left( \left| \tr{U(C)}\right|^2-1\right)\right\rangle
\label{adjloop}
\ee
which are invariant under the $Z_N$. We use the normalization that $W_A\sim 1$
in the naive continuum limit.

For the sake of simplicity let us postpone for a moment discussing
properties of the adjoint Wilson loop for the model \rf{partition} and review
some results~\cite{KhM81} for the case of the pure adjoint single-plaquette
lattice action
\be
S_A = -\frac{\beta_A}{2} \sum_\Box \left|\tr{U(\Box)}\right|^2
\label{adjaction}
\ee
where $\beta_A\sim1$ as $N\ra\infty$ in order to have a nontrivial large-$N$
limit. Due to the factorization at large $N$, the following formula holds
\be
W_A(C)=\left\langle \frac 1N \tr{U(C)}\right\rangle^2 + O(N^{-2})
\label{factorization}
\ee
where the average of the Wilson loop in the fundamental representation on the
r.h.s.\ should be calculated for the Wilson action
\be
S_{Wilson}=-N\bb \sum_\Box \Re \tr{U(\Box)}
\ee
with the coupling $\bb$ given by the self-consistency equation
\be
\bb=\beta_A W(\Box;\bb)
\label{selfconsistency}
\ee
where $W(\Box;\bb)$ is the plaquette average for the Wilson action
(with the coupling being $\bb$). When \eq{selfconsistency} possesses a
nontrivial solution which is valid in the weak coupling region so that
asymptotically
\be
\bb\ra \beta_A-\frac 14 \hbox{ \ \ \ as \ \ }\beta_A\ra\infty,
\label{bb}
\ee
one sees from
\eq{factorization} that the adjoint Wilson loop display the area law at
$N=\infty$ with the adjoint string tension being
\be
K_A=2K_{Fundamental}.
\ee
The perimeter law which is expected for the adjoint Wilson loop at finite $N$
enters the term of order $O(N^{-2})$.

Let us now return to the model \rf{partition}. Our idea is to define at
$N=\infty$ the `fundamental' Wilson loop, which enters, say, the correlator of
electromagnetic currents represented via sum over paths
or determines the string tension, by
\eq{factorization} taking the square root of the adjoint Wilson loop. This
procedure is unambiguous since the imaginary part of the fundamental Wilson
loop never shows up to the leading order of $1/N$-expansion.
It is crucial for this procedure the large-$N$ phase transition to occur
{\it before} the continuum limit sets in. Only in this case the induced
continuum theory would possess normal area law while otherwise one gets local
confinement.

Thus, we can define the set of QCD observables at the kinematical level despite
the $Z_N$-symmetry while this procedure works only at $N=\infty$.
It is a dynamical question whether the large-$N$ phase transition makes
this construction sensible.

\newsection{Large-$N$ phase transition}

Let us start again from the single-plaquette action \rf{adjaction}.
Since the only solution of \eq{selfconsistency} in the strong coupling
expansion is $\bb=0$, the adjoint Wilson loop $W_A(C)$ vanishes in this region
according to \eq{factorization}. One can directly verify this result
estimating the order in $1/N$ of the strong coupling expansion
which gives $W_A(C)\sim1/N^2$ for the action \rf{adjaction}. Therefore, as
we concluded in Ref.~\cite{KhM81}, there should be a first order phase
transition for the action \rf{adjaction} at some value $\beta_A^*<2$.

This phase transition was observed by the Monte--Carlo simulations for
$N=2\div6$ \cite{MC} with the critical value of $\beta_A^*$ growing from $1.6$
for $N=3$ to $1.8$ for $N=6$ what is close to the estimate $\beta_A^*<2$.
More detailed studies of \eq{selfconsistency} were performed in
Ref.~\cite{Sam83}. An uncertainty of the value of $\beta_A^*$ is related,
in particular, to an ambiguity of criterion
for the first order phase transition at large $N$.
The standard criterion stating that the phase transition occurs when free
energies of two phases coincide (the `Maxwell rule') may not work at
$N=\infty$ because the free energy itself is $\sim N^2$ and a barrier which
separates the two phases becomes infinite in this case. An alternative
criterion says that the phase transition is associated with the
point where a metastable weak coupling solution terminates.
Using the `Maxwell rule', Samuel obtained $\beta_A^*=1.54$ \cite{Sam83}.
Analogous studies of this phase transition using
mean field / variational technique~\cite{MF} yields $\beta_A^*=2.8$ while
the criterion based on terminating the metastable region gives
$\beta_A^*=1.7$ in better agreement with the Monte--Carlo data.
The large-$N$ phase transition is not related to breaking of a symmetry
and was shown to be associated with dynamics of
$Z_N$-monopoles which are condensed in the strong coupling phase
\cite{Monopole}.

An occurrence of the analogous phase transition for the Kazakov--Migdal
model has been advocated by Kogan, Semenoff and Weiss~\cite{KSW92}. We
support the consideration of this paper by the following argument. Let us
estimate the order in $1/N$ of the adjoint Wilson loops \rf{adjloop} in the
strong coupling expansion of the model \rf{partition} with the quadratic
potential $V[\Phi]$ using the induced action \cite{KM92} \be
S_{ind}[U]=-\frac 12 \sum_{\Gamma}
\frac{|\tr{U(\Gamma)}|^2}{l(\Gamma)m_0^{2l(\Gamma)}}.  \label{ind} \ee It is
easy to see now using the standard lattice technique that  $W_A(C)\sim
1/N^2$ in the strong coupling expansion while it is of order 1 in the naive
continuum limit. This estimate is quite similar to the one discussed in the
previous section for the case of the single-plaquette adjoint action. Thus,
we conclude the model undergoes a first order phase transition with
decreasing $m_0^2$.

Let us now estimate the location of this phase transition.
Following the scenario of Ref.~\cite{KM92},
we start from the induced action~\rf{ind} and decrease $m_0^2$.  A very
interesting question is whether the large-$N$ phase transition occurs before
the point \mbox{$m_0^2=D$} after which the quadratic action in
\eq{partition} is not bounded from below.
The point is that the action~\rf{ind} is not,
say, the classical action for an external field problem. One should
integrate $\hbox{exp}(-S_{ind}[U])$ over $U_\mu(x)$ with the Haar measure
according to the definition~\rf{partition}.

As $m_0^2\ra\infty$, typical configurations of $U_\mu(x)$ are uniformly
distributed on the group being independent on each link of the lattice.
This disordering suppresses the contribution of long loops to the sum over
paths on the r.h.s.\ of \eq{ind} --- the longer the loop the stronger the
suppression.  In usual lattice gauge theory the $U$-matrices become ordered
under the way to the continuum limit. The ordering is, however, only for
distances smaller than the correlation length while for large distances the
disorder is  needed for confinement. This ordering occurs either gradually
or is enhanced by the presence of a first-order phase transition.  It is,
however, a dynamical question whether this phase transition occurs in the
region $m_0^2\geq D$.

A simplest logical possibility for the model~\rf{partition} might be that
the large-$N$ phase transition occurs at some value of $m_0^2>D$ which is
large enough in order that the action \rf{ind} can be approximated by the
single-plaquette term, \ie coincides with the action \rf{adjaction} with
\be
\beta_A=\frac{1}{4m_0^8}.
\label{m8}
\ee
This would mean that the Kazakov--Migdal model were induce a lattice gauge
theory with the action~\rf{adjaction}.
Substituting the above numerical value $\beta_A^*\approx 2$, one would get
$m^2_*\approx 0.6$ which is too small. The next terms on the r.h.s. of \eq{ind}
are essential for such $m_0^2$ so that this situation is excluded by dynamics.

\newsection{The mean field analyses}

To study the large-$N$ phase transition, we apply
to our problem the mean field method,
which usually works pretty well for first order phase transitions,
quite analogously it was applied to
the single-plaquette adjoint action in Ref.~\cite{MF}.

To construct the mean field, we use the variational approach which was
advocated in the context of modern field theory by Sakita~\cite{Sak81}.
Let us introduce the trial partition function
\be
Z_0= \int \prod_{x,\mu} dU_\mu(x)
\e^{\frac{b_A}{2}\sum_{x,\mu} \left| \tr{U_\mu(x)} \right|^2 }
\label{trial}
\ee
which is a product of one-link integrals. We have chosen for them a simplest
form which possesses the $Z_N$ symmetry. The Jensen's inequality yields then
the following bound on the partition function \rf{partition}:
\be
Z \geq Z_0 \int \prod_{x}d\Phi(x) \e^{-\sum_x N\tr{V[\Phi(x)]}
+\left\langle  \sum_{x,\mu}\left( N\tr{\Phi(x)U_\mu(x) \Phi(x+\mu)
U_\mu^\dagger(x)}- \frac{b_A}{2} \left| \tr{U_\mu(x)} \right|^2
\right) \right\rangle_0}\;,
\label{bound}
\ee
where $\langle \ldots \rangle_0$ means averaging w.r.t.\ the trial action.
Since the exponent contains the sum of one-link averages, it can be
expressed via the one-matrix integral
\be
\h^2 = \frac{\int dU
\e^{\frac{b_A}{2}\left|\tr{U} \right|^2} \frac{1}{N^2}\left|\tr{U} \right|^2 }
{\int dU \e^{\frac{b_A}{2}\left|\tr{U} \right|^2}}\;,
\label{mf}
\ee
by the formula
\be
\left\langle
\tr{\Phi(x)U_\mu(x)\Phi(x+\mu)U_\mu^\dagger(x)}
\right\rangle_0=\h^2 \tr{\Phi(x)\Phi(x+\mu)}\;.
\ee

The idea of the variational mean field method is to fix $b_A$ from the
condition that the trial ansatz \rf{trial} would give the best approximation
to $Z$ in the given class. Calculating the derivative w.r.t.\ $b_A$ and
taking into account that $\h$ depends on $b_A$ according to \eq{mf}, one
finds the maximum of the r.h.s.\ of \eq{bound} at
\be
b_A= \frac{\int \prod_{x}d\Phi(x)
\e^{\sum_x N\tr{\left(-V[\Phi(x)]
+\h^2\sum_\mu \Phi(x)\Phi(x+\mu)\right)}}
\fr 1N \tr{\Phi(0)}\Phi(0+\mu)}
{\int \prod_{x}d\Phi(x)
\e^{\sum_x N\tr{\left(-V[\Phi(x)]
+\h^2\sum_\mu \Phi(x)\Phi(x+\mu)\right)}}}\;.
\label{b_A}
\ee
This is the final formula that relates $b_A$ to the potential $V[\Phi]$
providing the r.h.s.\ of \eq{mf} is known as a function of $b_A$.
Its standard interpretation is that one obtains the one matrix integral on
the r.h.s.\ of \eq{mf} replacing the matrix $[U_\mu(x)]_{ij}$ in the
partition function $\rf{partition}$ by the mean field value $\h \delta_{ij}$
on each link of the lattice except the given one $(0,0+\mu)$ while \eq{mf}
gives a self-consistency condition at this link.

The one-matrix integral on the r.h.s.\ of
\eq{mf} was first calculated by Chen, Tan and Zheng~\cite{MF}
(see also Ref.~\cite{Sam83}). In the weak coupling region where
$b_A>2$ or $1/2 \leq \h \leq 1$, the result can be represented as
\be
2(\h-\h^2)=\frac{1}{b_A}\;.
\label{wc}
\ee
We have written the self-consistency condition in the form which would be
convenient to study the phase transition.

Let us consider the case of the quadratic potential $V[\Phi]$ when the mean
field analyses is drastically simplified. The gaussian integral on the
r.h.s.\ of \eq{b_A} can easily be calculated to give
\be
b_A=\frac{1}{\h^2}\int_0^\infty d\alpha
\,\e^{-\frac{\alpha m_0^2}{\h^2}} \; \hbox{I\/}_0^{D-1}(\alpha)
\; \hbox{I\/}_1(\alpha)
\label{tildeb}
\ee
where \/I$(\alpha)$ are modified Bessel function.

Eqs. \rf{mf}, \rf{tildeb}  can be analyzed similarly to those of
Ref.~\cite{MF}.
The r.h.s.\ of \eq{tildeb} multiplied by
$\h^2$ monotonically increases with decreasing $m_0^2/\h^2$ with maximal
value $\approx 0.1$  at $m_0^2/\h^2=4$ and then diverges. This value is too
small to provide a solution to \eq{wc} with $\h>1/2$ as it should be for the
weak coupling solution.

We conclude, therefore, that for the quadratic potential $V[\Phi]$ {\it there
is no first order large-$N$ phase transition} for $m_0^2\geq 4$ when the
model is well-defined.
Notice that we exclude a possibility that the large-$N$ phase transition
occurs exactly at $m_0^2=4$. According to \eq{wc} this would occur if
$b_A$ were $\geq 1/2$ at this point which is not the case.

Let us mention that the gaussian integral over $\Phi$ on the r.h.s.\ of
\eq{b_A} for the quadratic $V[\Phi]$ can be represented in the $1/m_0^2$
expansion as the sum over paths:
\be
b_A=\frac{1}{m_0^2}\sum_{\Gamma_{x,x+\mu}}
\left(\frac{\h}{m_0}\right)^{2l(\Gamma)}
\label{sum}
\ee
where the sum goes over the open loops with the endpoints $x$ and $x+\mu$.
This sum
is not exactly the same as what would appear if the mean field method were
applied directly to the action \rf{ind}.
The point is that if a link is passed back and forth,
it does not contribute to the r.h.s.\ of \eq{ind}
because $U$ is unitary while it is taken into account
on the r.h.s.\ of \eq{sum}. In particularly first few terms of the series
expansions of r.h.s.\ of \eq{tildeb} in $D=4$ read
\bea
b_A\h^2=  \fr 12
\left(\frac{\h}{m_0}\right)^4 + \fr {21}{8} \left(\frac{\h}{m_0}\right)^8 +
 20 \left(\frac{\h}{m_0}\right)^{12} +
 \fr {23765}{128} \left(\frac{\h}{m_0}\right)^{16} +
 \fr {124047}{64} \left(\frac{\h}{m_0}\right)^{20} \nonumber \\ +
 \fr {1397319}{64} \left(\frac{\h}{m_0}\right)^{24} +
 \fr {4148859}{16} \left(\frac{\h}{m_0}\right)^{28} +
 O \left(\left(\frac{\h}{m_0}\right)^{32}\right)
\label{series}
\eea
which differs to order $O(1/m_0^8)$ from what one would obtain from \eq{m8}.
All the extra terms are, however, positive so that $b_A$ can be considered
as an upper limit which is enough for our purposes.

\newsection{Discussion}

The main conclusion from the fact that the large-$N$ phase transition does
not occur for the model \rf{partition} with the quadratic potential
$V[\Phi]$ is that it remains in the strong coupling phase with
local confinement and can not induce QCD. This phase is pretty trivial:
the averages of all the Wilson loop vanish at $N=\infty$ except of those
with vanishing minimal area, in contrast to the strong coupling expansion of
the lattice gauge theory with the Wislon action%
\footnote{The existence of a master field for the Wilson action in the
strong coupling region was first advocated by Kazakov, Kozhamkulov and
Migdal~\cite{KKM87}.}.
Therefore, one should incorporate the self-interaction of $\Phi$ and look
for the large-$N$ phase transition. The mean field method which leads in
the case of an arbitrary potential $V[\Phi]$ to Eqs.~\rf{b_A} and \rf{wc}
can be used to estimate the location of the large-$N$ phase transition.

As we discussed already, the large-$N$ phase transition should occur {\it
before} the one associated with the continuum limit in order that the
Kazakov--Migdal model induce normally confining QCD. One might tempt to
relate the latter phase transition with the one separating confining and
Higgs phases which always occurs in gauge theories. A continuum theory with
confinement can be usually obtained by approaching this phase transition
from above while approaching from below one gets the deconfining Higgs
phase. If one accepts this scenario, the main problem with inducing QCD by the
Kazakov--Migdal model is to study whether the large-$N$ phase transition is
indeed separated from the Higgs one. If the two phase transitions coincide,
this would mean that one passes from the phase with local confinement
directly to the Higgs phase while the phase with normal confinement were
missing. Monte--Carlo simulations of the Kazakov--Migdal model might help to
answer this question.

The mean field method could be useful to study the Higgs phase
transition as well. In the Higgs phase one would get a condensate of the
$\Phi$-field, $\Phi_*$, with a nonsymmetric distribution of eigenvalues of
$\Phi_*$ which violates the $SU(N)$ quite similarly, say, to the Higgs phase
transition in the Georgi--Glashow model.
One should add then one more self-consistency condition to determine
$\Phi_*$. Substituting $\Phi(x)$ by
an (independent on $x$) mean-field value $\Phi_*$ for all sites of the
lattice except given one, we get the self-consistency condition
\be
\fr 1N \tr{\frac{1}{\l-\Phi_*}}=\frac{\int d\Phi
\e^{N\tr{(-V[\Phi]+D\h^2\Phi\Phi_*)}}
\fr 1N \tr{\frac{1}{\l-\Phi}} }
{\int d\Phi \e^{N\tr{(-V[\Phi]+D\h^2\Phi\Phi_*)}}}\;.
\label{phi}
\ee
This condition means that the saddle-point configuration of the integral
over $\Phi$ coincides with $\Phi_*$ and $\l$ plays the role of a spectral
parameter.

The matrix integral in \eq{phi} coincides with the partition function of the
hermitian one-matrix model in an external field, $\Phi_*$.  While its
solution for a quartic potential $V[\Phi]$, which has been taken into account
in Ref.~\cite{Mig92a}, is not yet known, this model was explicitly solved in
genus zero (the $N=\infty$ limit) for a cubic potential~\cite{Konts} and for
a logarithmic potential~\cite{CM92}. We hope the obtained results could be
useful in this context.

\section*{Acknowledgments}

Yu.M. thanks A.Migdal, A.Morozov, G.Semenoff and N.Weiss for e-mail
correspondences.

\section*{Added note}

When this paper was being prepared for publication, there appeared more
papers~\cite{new} on the Kazakov--Migdal model. The results by Gross
agree with ours for the quadratic potential while the Monte--Carlo study
by Gocksch and Shen seems to indicate that for $N=2$ the `large-$N$' phase
transition coincides with the Higgs one.

\eop

\end{document}